\pdfoutput=1
\documentclass[conference]{IEEEtran}
\usepackage{cite}
\usepackage{amsmath,amssymb,amsfonts}
\usepackage{algorithmic}
\usepackage{graphicx}
\usepackage{textcomp}
\usepackage{xcolor}
\usepackage{hyperref}

\def\BibTeX{{\rm B\kern-.05em{\sc i\kern-.025em b}\kern-.08em
    T\kern-.1667em\lower.7ex\hbox{E}\kern-.125emX}}

\graphicspath{{./Images/}}
\DeclareGraphicsExtensions{.pdf, .jpg, .png}
\begin{document}

\title{A Highly Configurable Hardware/Software Stack \\for DNN Inference Acceleration}

\author{
\IEEEauthorblockN{Suvadeep Banerjee, Steve Burns, Pasquale Cocchini, Abhijit Davare,\\
Shweta Jain, Desmond Kirkpatrick, Anton Sorokin, Jin Yang, Zhenkun Yang}
\IEEEauthorblockA{\textit{Intel Labs}}
}

\maketitle

\begin{abstract}
This work focuses on an efficient design methodology for domain-specific accelerators.
We employ an Agile development approach, with feature-by-feature enhancement of a vertical
development stack. This development methodology has been applied to the TVM/VTA inference
accelerator. Along the way, we have enhanced the VTA design space and enabled end-to-end
support for additional workloads. This has been accomplished by augmenting the VTA
micro-architecture and instruction set architecture (ISA), as well as by enhancing the
TVM compilation stack to support a wide range of VTA configurations.

The VTA \textit{tsim} implementation (CHISEL-based) has been enhanced with fully pipelined versions
of the ALU and GEMM execution units. In \textit{tsim}, memory width is now parameterized to range
between 8-64 bytes per cycle. Field widths and ISA encoding have been made more flexible
across multiple targets to support larger addressable scratchpads. A handful of new instructions
have been added to enable new or faster functionality. These include: element-wise 8-bit
multiplication to support depthwise convolution, load with a choice of pad values to support
max pooling, and a clip instruction to support faster execution of a common pattern in ResNets.
Support for additional layers, enhanced double buffering allowing for greater scratchpad
utilization, and runtime enhancements to lower uop count have also been added.

A significant increase in performance is seen for the \textit{tsim} target just using the fully pipelined
versions of ALU and GEMM: $\sim$4.9x fewer cycles with minimal area increase to
run ResNet-18 under the default configuration. By varying existing and newly added parameters,
configurations featuring a further $\sim$11.5x decrease in cycle count at a cost of
$\sim$12x greater area can be instantiated. Tens of intermediate points on the
area-performance pareto curve are shown, showcasing the balance of execution unit sizing, memory
interface width, and scratchpad sizing which is required to extract good performance from this
relatively simple micro-architecture. Finally, VTA is now able to run Mobilenet 1.0 and all
layers for ResNets, including the previously disabled pooling and fully connected layers.

The TVM/VTA architecture has always featured end-to-end workload evaluation on RTL in a matter of
minutes. With our modifications, it now offers a much greater number of feasible
configurations with a wide range of cost vs. performance. All capabilities mentioned here, and
more, are available in open-source forks of the `tvm' and `tvm-vta' repositories. A subset of
these capabilities have already been upstreamed.

\end{abstract}


\section{Introduction}

\subsection{Motivation}

Our motivation for exploring specialization of hardware (accelerators) to increase compute performance comes from some of the early ``dark silicon'' work \cite{hadi2011} \cite{fourhorsemen2012}, where the authors point out that while power has limited frequency scaling of general purpose compute, we have the opportunity to spend transistors on specialized processing for specific applications. Yet, the cost of developing new accelerators is too high. Moore's law, which states that the cost per transistor decreases by 2x every process generation, does not help here unless the target accelerator has a very large market. For smaller markets, non-recurring engineering (NRE) costs dominate and any reduction of these costs could proportionally reduce per unit costs. Concentrating on new design methodologies that reduce NRE is therefore an important research direction.
Hennessey and Patterson, in their 2018 Turing award paper \cite{hennesseypatterson2019}, discuss this in their call for the development of Domain-Specific Accelerators.

Both software and hardware systems incur substantial NRE costs, as well as design and verification aspects of both. Developing design methodologies that jointly optimize both software and hardware is a focus of this work. In looking around for a domain to make our methodology research more concrete, we decided to explore inference accelerators for Machine Learning (ML), partly because of our initial exposure to DARPA's RTML (Realtime ML) BAA \cite{RTMLBAA} and early work with a performer on the project. In studying existing solutions to the problem, the TVM software stack and the VTA hardware accelerator looked promising as a good starting point for the methodology research we wanted to explore.

In looking at the existing solution, we found a fairly well developed software stack (TVM \cite{tvm2018}) and an immature hardware implementation (VTA \cite{moreau2018vta}). The hardware description was written in Chisel\cite{chisel_dac2012} which our team had existing experience with. However, it was incomplete in a variety of ways and did not compute the correct answers on at least one of the targeted FPGA platforms.
So we decided to build up a development methodology to improve TVM/VTA so that it was as good as the underlying architecture would allow it to be. We wanted to show this with different amounts of hardware resources as well. This, in the end, involved correcting several things: 1) lack of parametrization, 2) poor end-to-end performance, 3) inflexible memory systems, 4) poor instruction sequences generated by the compiler (e.g., double buffering) and 5) non-optimal scratchpad sizes. To do all of this, we need to develop design methodologies along the way, including: 1) a test driven development (TDD) solution of developing hardware blocks, 2) visualization techniques for understanding the performance of the system, 3) verification flows to determine where hardware implementation state diverges from software simulator state, 4) techniques to improve the compiler, 5) a continuous integration (CI) system to ensure software simulation, RTL simulation and FPGA simulation remained correct on each check-in, and 5) flows and methods for performing physical design as an ASIC.

\subsection{Related Work}

An early Domain Specific Language (DSL), Halide \cite{halide2013}, introduced
the important concept of separating specification of computation from the
scheduling, which is dependent on the hardware target, an important advance in
targeting workloads flexibly to different types of hardware acceleration. Recently, Deep
Neural Networks have received great attention in DNN compilation stacks such as
Tensorflow\cite{tensorflow2016}, PyTorch \cite{pytorch2019}, MXNet
\cite{mxnet2015} which map the deep learning operators to vendor-provided
kernel libraries for execution on CPU and GPU hardware with high performance.
Compiler frameworks such as Relay
\cite{Roesch_2018} and MLIR \cite{lattner2020mlir} have evolved for lowering the deep learning computations into these kernel libraries. Due to the high manual engineering
effort required to develop and tune these libraries, there has been a recent
proliferation of deep learning compilers that can automatically optimize the
low-level implementation of tensor operators for a given accelerator
design. One class are polymorphic compilers that pose program optimization as
an integer linear programming problem, such as Tiramisu \cite{tiramisu2019},
FlexTensor \cite{flextensor2020}, and Tensor Comprehensions
\cite{comprehensions2018}). Another class are those based on graph
optimization techniques \cite{graph1, graph2} and auto-tuners such as Relay
\cite{Roesch_2018}, OpenTuner \cite{opentuner2014}, NeuroVectorizer
\cite{neurovectorizer2020} and AutoPhase \cite{autophase2020} that achieve a
certain level of auto-tuning. A final class, using a variety of schedule
optimization techniques are Halide with its auto-schedulers \cite{halide1,
  halide2, halide3}, TVM \cite{tvm2018} with AutoTVM and Ansor
\cite{ansor2020}.

DNNWeaver2 \cite{dnnweaver_micro16} is an early work in exploring software/hardware co-design in the DNN inference domain, serving as an inspiration for
this work.  On the hardware side we see innovations in flexible hardware
construction languages that help generate new accelerators, like Chisel, PyMtl3 \cite{pymtl3_2020}, Magma \cite{magma_DAC2020}.
\cite{tvm_vta2018} has extended the targets of TVM to include a special
inference hardware engine called Versatile Tensor Accelerator (VTA). Our effort
is an extension of this TVM/VTA compiler/hardware stack.

\section{TVM/VTA Stack Overview}

\subsection{Versatile Tensor Accelerator Microarchitecture}

\begin{figure}
  \includegraphics[width=\linewidth]{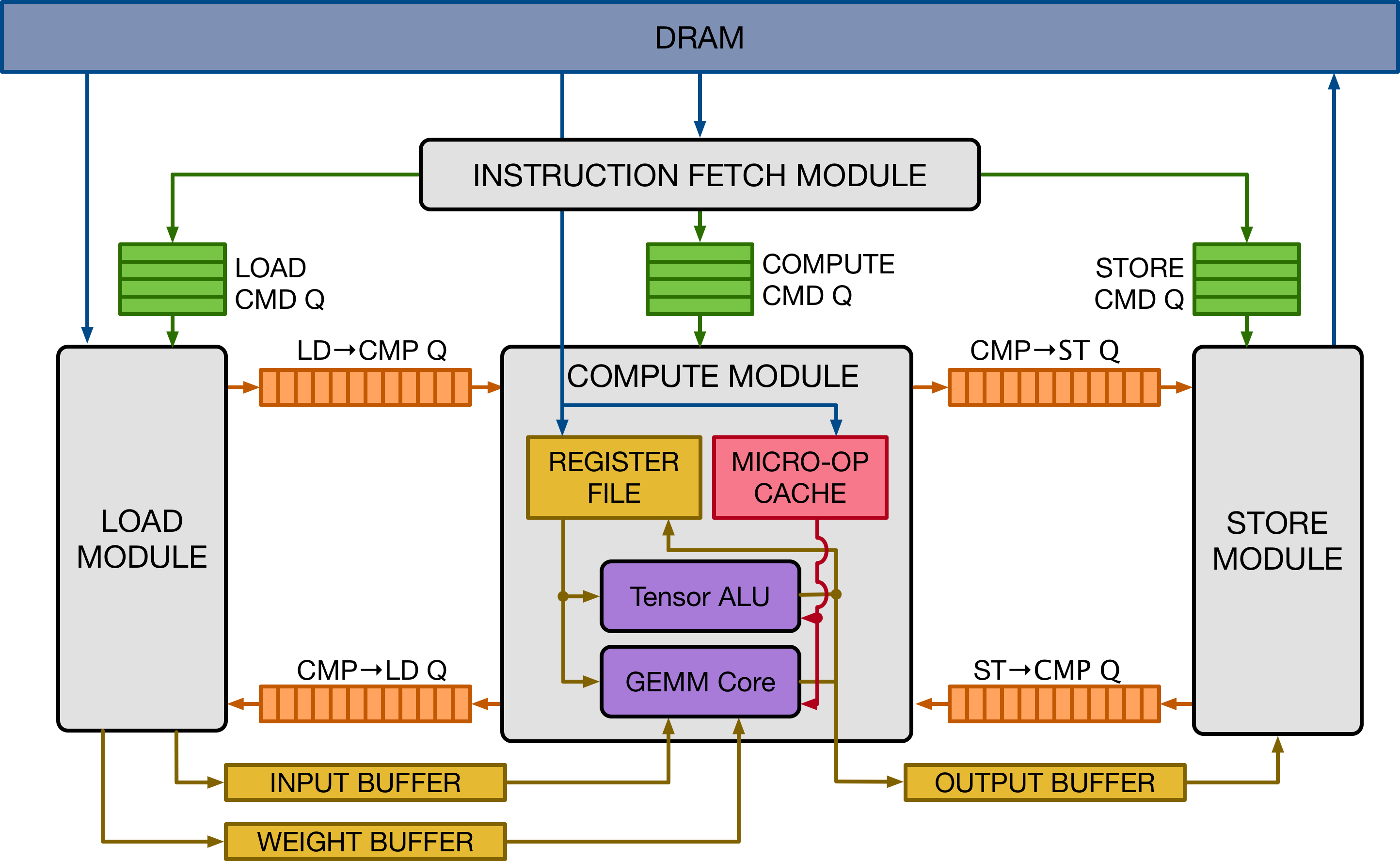}
  \caption{VTA Microarchitecture \cite{moreau2019hardware}}
  \label{fig:vta_uarch}
\end{figure}

Figure~\ref{fig:vta_uarch} illustrates the Versatile Tensor Accelerator (VTA)
hardware, a parameterized load-compute-store micro-architecture comprising
fetch, load, compute, and store modules which use command queues to propagate
inference compute tasks through the machine, and using scratch buffers to hold
input, weight, and output (accumulate) data. The fetch module loads inference
task micro-operation instructions from main memory and dispatches command
sequences to the load, compute and store modules. The load module performs
strided memory access to load input, weight, and bias tensor tiles from DRAM
into their on-chip buffers. The compute module contains a GEMM core for
processing convolutional layers and an ALU for activation, normalization, and
pooling layers. The GEMM unit as published has pipeline depth of 4 (single
operand) or 5 (2 operands), the machine only can issue another GEMM instruction
after completion resulting in an initiation interval (II) of 4. The ALU as
published is not pipelined. Finally, the store module transfers the contents of
the accumulate buffer to main memory upon completion of the inference task.

The load, compute, and store modules all obtain instructions from separate
queues, and can execute in parallel. However, instructions within these queues
may have read/write dependencies with instructions from other queues. Such
dependencies are added by the compiler and enforced through the usage of the 4
dependency queues shown in Figure~\ref{fig:vta_uarch}: \textit{LD}-$>$\textit{CMP}, \textit{CMP}-$>$\textit{LD},
\textit{CMP}-$>$\textit{ST}, and \textit{ST}-$>$\textit{CMP}. The compiler sets a subset of 4 dependency bits in
each instruction: \textit{pop prev}, \textit{pop next}, \textit{push prev}, and \textit{push next}. Here, \textit{prev}
and \textit{next} refer to the queues on the left and right sides, respectively, of
the module which executes the instruction. \textit{pop} commands block until a token
is present in that queue before executing the instruction, while \textit{push}
commands insert a token into that queue. In essence, these 4 bits force
sequential execution only when needed to avoid race conditions on the 3 shared
resources which are touched by the modules: the input, weight, and output
buffers. Setting extraneous dependency bits can result in longer cycle counts
or even deadlock.

The VTA micro-architecture is modeled in a bit-accurate manner in two ways:
\textit{fsim}, which is a behavioral description of the machine and Chisel, which is a
parameterized RTL description. The fundamental parameters of the machine are
the BLOCK\_IN and BATCH size of the input and BLOCK\_OUT of the weight and
accumulate tensors, as well as the size of the input, weight, and accumulate
buffers. As published, the VTA hardware presumes a 64-bit databus for memory
access, and the working Chisel configuration was limited to
BLOCK\_IN=BLOCK\_OUT=16 with BATCH=1. The available ALU operations limit VTA
operation to most, but not all, layers of ResNet inference workloads.  Yet the
RTL model is small enough to execute these workloads in minutes of simulation
time, making it an interesting target for design space exploration.

\subsection{Instruction Set}

Instructions and micro-operations (uops) are used to program the VTA accelerator, and both were extended in order to
allow greater microarchitectural flexibility. New instructions and variants were also added to enable additional
functionality.

The VTA ISA contains 5 instructions: GEMM, ALU, LOAD, STORE, and FINISH. The GEMM and ALU instructions are 
\emph{compute} instructions and each is associated with a sequence of uops. Each uop in the sequence holds base 
addresses for the scratchpads that are read/written within the loops which characterize the instruction. The 
LOAD and STORE instructions are capable of being executed in parallel with \emph{compute} instructions, and are 
not associated with uops. By default, each instruction is 128 bits wide while each uop is 32 bits wide.

Instructions and uops are emitted by the runtime, while they are consumed by each of the hardware targets 
(\emph{fsim}, \emph{tsim}, etc). Both sides need to have a common understanding of the structure and semantics,
while hardware needs to instantiate the structures (e.g. queues) and decoders which are required to process the 
instructions and uops. Due to the two-level breakdown of VTA into a runtime along with hardware, the
upper compiler/runtime interface can remain more insulated from ISA changes.

Our goals to change the shapes of tensors processed by GEMM and ALU instructions and modify the sizes of scratchpads
naturally result in field width changes within both instructions and uops.

For uops, we also extended the size of uops since not enough spare bits were available. Wider uops can support wider
fields, allowing larger scratchpads, but also require additional storage and memory bandwidth.

For instructions, we retained the 128-bit width as a constant. Permitting increased width for certain fields uses up
spare bits. After exhausting available spare bits, we resorted to shrinking other field widths in order to fit within the
instruction width constraint.

A JSON configuration file is the only compile-time construct consumed by the compiler, runtime, as well as all
hardware targets. Variable field widths can be captured in the JSON file. New JSON parameters need to be handled
across multiple languages. Compile-time checks - such as ensuring instruction width constraints are not violated -
need to be implemented as well.

\subsection{Compiler}
\label{sec:compiler}

The software stack for targeting the VTA hardware starts with ingesting deep learning models expressed in frameworks such as Pytorch\cite{pytorch2019}, TensorFlow\cite{tensorflow2016} and MxNet\cite{mxnet2015} into Relay\cite{Roesch_2018} IR (Intermediate Representation) for representing computational graphs. The TVM compiler decomposes the workload into appropriate VTA-compatible operator schedules where the operators are further partitioned by loop manipulations into VTA hardware intrinsics such as GEMM/ALU or DMA load/store. The VTA compilation stack includes a software defined runtime that performs JIT compilation of the TVM compiler generated API calls for creating the final instruction sequence. 

For extracting high performance out of VTA, the TVM compiler utilizes the decoupled access-execute philosophy \cite{dae1982} by exploiting thread parallelism. The virtual thread injection pass of the TVM compilation stack partitions schedule tasks into different execution contexts such that load, store and compute operations can concurrently execute. This double buffering is effective in hiding the memory latency such that one half of the scratchpad is used by DMA load/store, where the compute module operates on the other half. The compiler manages this fine-grained parallelism by analyzing subsequent load, compute and store nodes in the IR to determine the local buffer addresses being used. This information is used to create the memory access and compute instruction JIT calls with requisite \textit{push} and \textit{pop} tokens injected into these calls.

The SW stack implements a SW-defined runtime system that manages heterogeneous execution of workloads on CPU and VTA. Instead of the TVM compiler directly generating code for the VTA target, the runtime exposes a high-level API of hardware intrinsic calls that the compiler can lower schedules onto. The JIT runtime creates the instruction stream from the compiler generated recipe by managing DRAM load/store and managing micro-kernel generation. The runtime interprets the \textit{push} and \textit{pop} dependency tokens inserted by the compiler between instruction calls to control the sequential execution of the instructions. The flexibility of the JIT runtime allows layers of a deep network to be either executed on the CPU or offloaded to the VTA, thus ensuring that a DNN can be executed on VTA even if the accelerator doesn't support all layers.

\section{Development Methodology}

In this section, we provide an overview of our Agile development methodology: how we decide to implement a
new incremental feature and how we ensure it is implemented correctly.

\subsection{Performance Analysis and Visualization}

\begin{figure}
  \includegraphics[width=\linewidth]{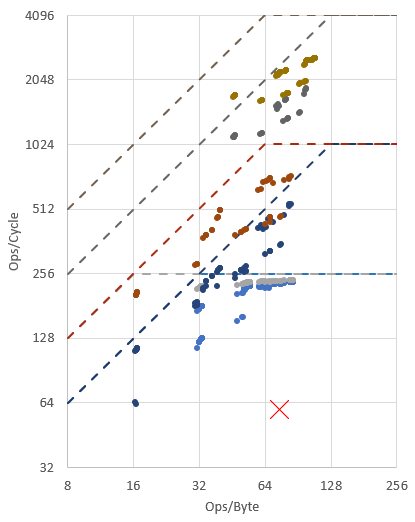}
  \caption{Roofline Chart: This is a log-log chart with Ops/Byte on the x-axis and Ops/Cycle on the y-axis. The horizontal dashed lines represent compute bounds based on the number of simultaneously operable compute units. The diagonal dashed lines correspond to memory bandwidth limit (the intercept with the vertical line Ops/Byte = 8 corresponds to the bandwidth in Bits/Cycle).}
  \label{fig:roofline}
\end{figure}

\begin{figure*}
  \includegraphics[width=\textwidth]{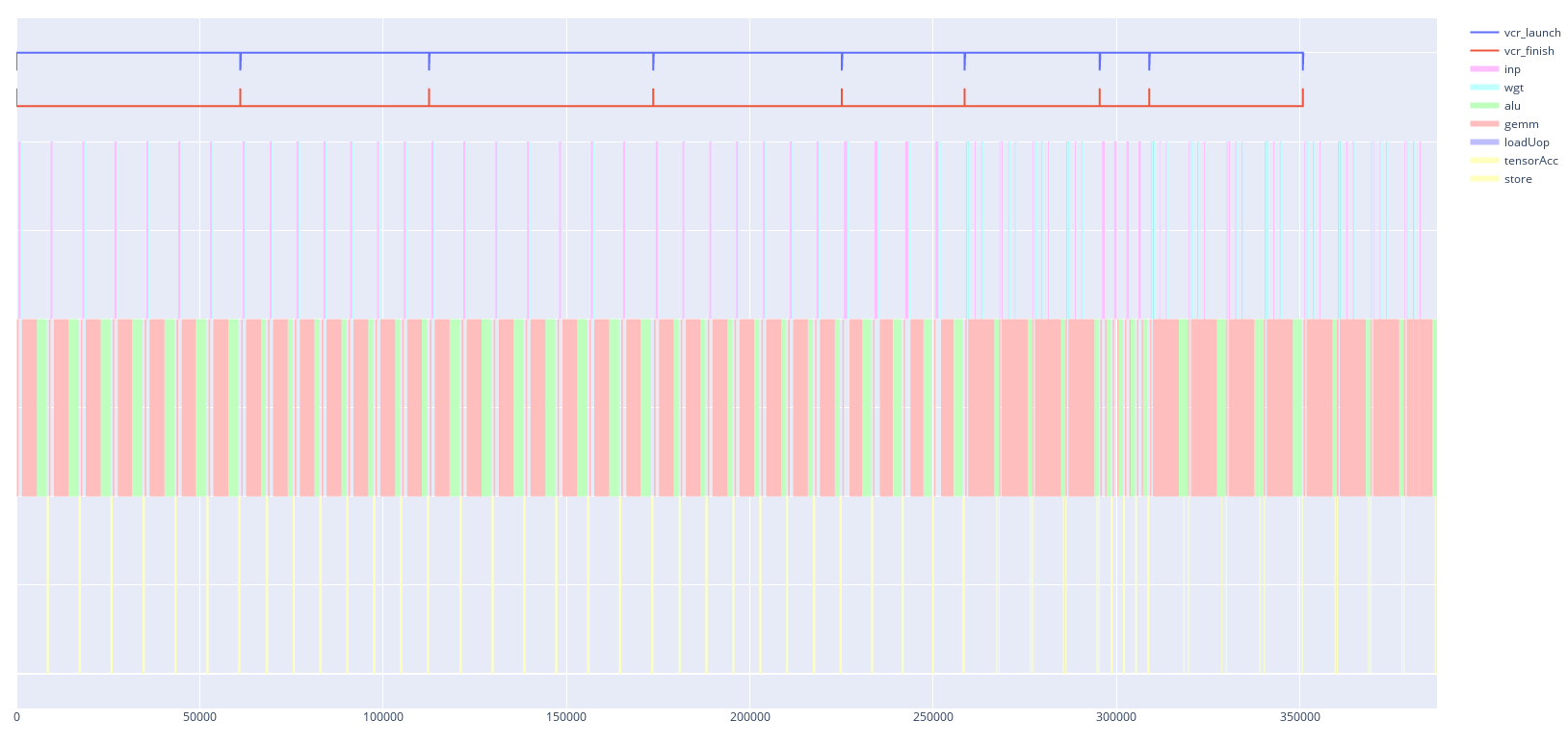}
  \caption{Process Utilization Visualization for a complete ResNet-18 workload. The three bars correspond to the {\em load},  {\em compute} and  {\em store} processes. This computation is compute bound because both {\em load} and {\em store} are idle for significant amounts of time. The red sections of {\em compute} correspond to GEMM activity and the green sections to ALU activity.}
  \label{fig:utilization}
\end{figure*}

\begin{figure*}
  \includegraphics[width=\textwidth]{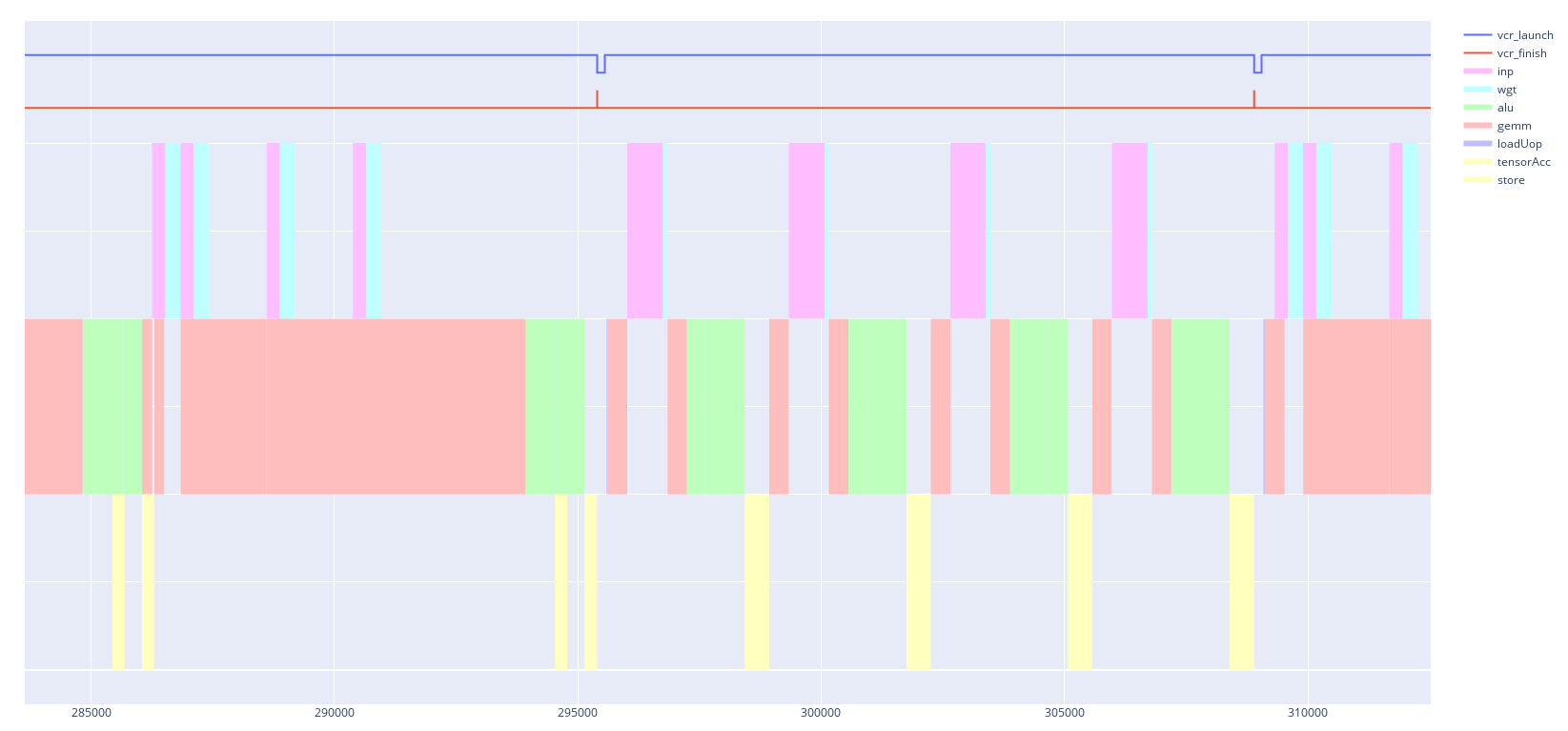}
  \caption{Process Utilization Visualization for three partial layers of the ResNet-18 workload. The complete layer (between the red ticks, {\em vcr\_finish}) shows a sequential ordering between the {\em load}, {\em compute} and {\em store} activities. This layer could likely be improved by double bufffering, allowing, for example, {\em load} and {\em compute} activities to run concurrently. The previous layer (to the left of the first red tick), shows this overlap and is likely implemented using double buffering.}
  \label{fig:zoomutilization}
\end{figure*}

Given an existing, mostly working TVM/VTA repository, we now had the problem of analyzing the existing
performance and determining what simple changes could be made to improve the design. We created two simple
visualizations to help with this analysis: 1) a roofline chart \cite{roofline2009} which can show in a single
diagram how closely a computation fully utilizes the available compute resources and memory bandwidth; and
2) an activity visualization of the three loosely-coupled processes in VTA ({\em load}, {\em compute}, and
{\em store}). Figure~\ref{fig:roofline} shows our Roofline chart for a variety of scratchpad sizes, number
of compute units, and memory bandwidths. Figure~\ref{fig:utilization} shows that all three processes can run
concurrently in a balanced design. Figure~\ref{fig:zoomutilization} show a more detailed view from which one
can easily see dependencies between the VTA processes.

\subsection{Always-alive Modeling}

The TVM/VTA inference stack spans many layers: workloads, compiler, runtime, behavioral model, and RTL. These
layers are implemented in a variety of languages. Development may be carried out by non-overlapping groups of
people. Feature enhancements to one layer may not be visible with end-to-end workload evaluation unless other
layers are modified in tandem. Conversely, unanticipated degradation in workload performance may occur due to
relatively minor changes in a given layer. Reducing the time required to see the effects of a given change
motivated the development of an always-alive model.

In a traditional `waterfall' development methodology, many features can be in flight simultaneously. The initial
decomposition of functionality into features is carried out at an early stage while the integration of features
takes place as one of the last stages during the development process. During feature development, implementations
are usually tested in isolation, for instance with unit tests written against the feature specification, in order
to facilitate parallel development. Since integration and end-to-end testing are deferred, information gleaned
from these activities cannot drive significant implementation changes without schedule slip.

In this work, we follow the Agile approach, where integration occurs frequently, giving fast insights into the
impact of a particular feature. This low-latency feedback requires automation of integration and
end-to-end evaluation. If manual effort were required, its aggregate cost across the development process would
be too high. At the beginning of development, it is important to bootstrap a baseline vertical stack which
exercises as much end-to-end behavior as possible. It is expected that these capabilities will be enhanced
and expanded during the development process: integration and evaluation should be incrementally
developed features.

For the TVM/VTA inference stack, end-to-end evaluation involves both functional and performance aspects. Running
inference on input image(s) and comparing the labels against expected results for both the behavioral and RTL
targets gives us an indication of functional correctness. Performance counters in the RTL model tracked over time
help us understand the performance impact of various features.

\subsection{Dynamic Trace-based Validation}

Central to any agile development methodology is the ability to quickly identify and correct any functional,
performance defects present in the design at the end of frequent integration runs. Indeed, without
such a capability, verification and validation tasks would quickly become a challenging bottleneck increasing
design turn-around-time to unacceptable levels.

To ensure a sufficiently high degree of functional coverage, a regression suite composed of several representative
end-to-end tests was setup to run during continuous integration cycles, with test content highly parameterized and
varying from a few instructions to entire inference workloads.

All tests were run for 4 supported targets of the VTA accelerator \cite{tvm_vta2018},
all sharing the same common runtime introduced earlier:
\begin{itemize}
    \item \textit{fsim}: C++ behavioral model. Low design complexity as compared to other targets.
    \item \textit{bsim}: HeteroCL \cite{heterocl} behavioral model.
    \item \textit{tsim}: Cycle accurate model simulating Chisel-generated RTL using Verilator \cite{verilator}.
    \item \textit{de10nano}: FPGA emulation model using the DE10-Nano board \cite{de10nano} (based on the 
    Intel Cyclone V FPGA), running a cycle accurate model synthesized from Chisel-generated RTL.
    \item \textit{pynq}: FPGA emulation target using the Pynq board \cite{pynq} (based on the Xilinx Zynq FPGA),
    running a VTA cycle accurate module synthesized from a C++ High Level Synthesis model using Vivado HLS \cite{vivado}.
\end{itemize}

When a test failed to produce the expected result for any given target, the test was rerun in \textit{trace mode}
under the same conditions and for all supported targets, producing a configurable dump of architectural states.
The trace produced by the failing target was then compared to the trace produced by another passing target.
A detailed comparison pinpointed the location in the trace where the behavior of the failing target
diverged from that of the passing target. The divergence point was then used to cross-reference the failing target
code and find the location of the defect.

The methodology required the instrumentation of arbitrary architectural states across behavioral and RTL
targets written in different languages, i.e. C++, Python, Scala. To this purpose, a trace manager module was 
developed in each language featuring a common interface that allowed for the unambiguous specification 
of the same architectural states to be monitored and traced.

For flexibility and ease of use, tracing was parameterized through the use of user selectable \textit{trace modes} 
allowing the generation of traces with different levels of granularity, both in time and across runtime and architectural
state spaces. For instance, generating traces for a specific scratchpad SRAM memory, at specific time events in
the GEMM or ALU units.

This methodology was found to be very effective at quickly locating defects especially when comparing the
\textit{tsim} cycle accurate RTL target versus the \textit{fsim} behavioral reference C++ target,
through a careful selection of relevant runtime and architectural states with invariant properties exhibited in
both models.

\subsection{Unit Testing}

Unit testing provides the finest granularity and fastest testing capabilities for test-driven RTL design.
It is based on Chisel Testers. Chisel Testers provide a library for functional testing
at the lowest design hierarchy level that gives fast turnaround time for a
cycle-accurate simulation. It also provides cycle-accurate traces for power and
performance analysis.  Chisel Iotesters PeekPokeTester is used to communicate with RTL.
Iotesters provide connection with three simulation backends. Treadle is a FIRRTL level simulator.
It provides performance comparable with verilator and has a very short spin-up
time. Treadle cannot handle black box Chisel modules which are provided as Verilog
sources; therefore, vcs or verilator should be used to run tests with memory
compiler hard IP blocks. Chisel Verilog compiler can replace memories with
external reference to macro memories if it is required for further physical
synthesis flow. A Chisel-based macro memory compiler is used to select and connect memory
compiler hard IP blocks and provide macro memory blocks for given memory configurations.
PeekPokeTester is driven by Scala test driver code that can communicate at different levels of VTA
hierarchy. The highest level of communication is VTAShell AXI interface level. Another level
is Core vme/vcr communication level. To drive both levels, a Scala code to
create instructions/opcodes/data is used. It is also used to encode execution
parallelism ordering. A GEMM multiplication with double buffering in scratchpad
and ALU load-operate-store with different padding can be executed at those levels
for all feasible configurations automatically. A multiple read/write requests
memory model with a constant latency is implemented. Tests created for original
configuration drive development and validation of a new functionality.
While the largest configurations of VTA could take several minutes to compile
Verilog from Chisel sources, it is more practical to run tests first on a smaller
VTA which can be compiled in less than a minute. The simulation time of most unit tests is less than a minute that puts checking of a small feature in
a category of almost interactive development.

\section{Results}

\subsection{Performance Analysis Driven Improvements}

We realized the main hardware related performances improvements in a greedy incremental way.
Performance analysis, in particular utilization charts like Figure~\ref{fig:utilization}, showed that the design was compute bound on the ResNet-18 workload.

\subsubsection{Pipelining the GEMM Unit}

Analysis showed that a significant amount of time was being spent in the GEMM unit. While the matrix-vector units were fully pipelined, they were only being used once every fourth cycle. The main state machine for the {\em compute} unit was sequencing through four states: {\em fetch uop}, {\em fetch data}, {\em MAC}, and {\em writeback} for each computation. Since the control flow is completely determined,
a very simple pipelining strategy transforms this to initiation interval (II) equal to 1. The basic idea is to have a single control unit generating a sequence of three dimensional loop indices. We then pipeline the UOP lookup, the weight and input scratchpad lookups, and finally the vector-matrix product calculation. A simple inflight queue is used to flush the pipeline after completion of each instruction (which typically takes hundreds of cycles to complete).

RTL changes were done locally to the {\em TensorGEMM} block using handcrafted unit tests following a TDD methodology. While these tests passed, full end-to-end integration tests (on some subset of the layers of Resnet-18) failed. The case of this failure was an address staging bug in another unit ({\em LoadUop}) which was uncovered now because uops are being fetched every cycle instead of once every four cycles. (This bug would have been easy to catch using the dynamic tracing facility, which was not available when this RTL work was performed). Careful study of signal traces was required to root cause and correct this error. Significant work went into developing a complete Chisel-based testbench of the {\em TensorGEMM} including not only tests on the sequence of indices generated, but also on the values produced by the matrix-vector units. This testbench included a Scala-based model of the {\em LoadUop} block [considered to be part of the environment to {\em TensorGEMM}] and was not helpful in finding the bug.

\subsubsection{Pipelining the ALU Unit}

With the GEMM unit no longer the bottleneck, we decided to similarly improve performance of the ALU which had an II of either 4 or 5 (depending on whether one or two operands were needed). Initial analysis suggested that little gain could be achieved by improving this unit that was rarely used. This analysis mistakenly assumed an II=1. Redoing it we discovered using the utilization diagrams shown in Figure~\ref{fig:utilization} that approximately 50\% improvement was possible at least for some of the ResNet-18 layers.

Again, RTL changes were done locally to the {\em TensorALU} block. This time, handcrafted unit tests following the TDD methodology were only developed for the index generation logic. The final pipeline was II=2 for two operand instructions (because the accumulator register file only allows one read access per cycle) and II=1 for one operand (immediate) instructions. Final testing was done at the integration-test level using the dynamic tracing capability to find some wiring errors at the datapath level. This hybrid approach (Chisel-based tests for control streams and dynamic tracing for data steering) is recommended as a good balance that utilizes the best of both testing methodologies.

\begin{figure}
  \includegraphics[width=\linewidth]{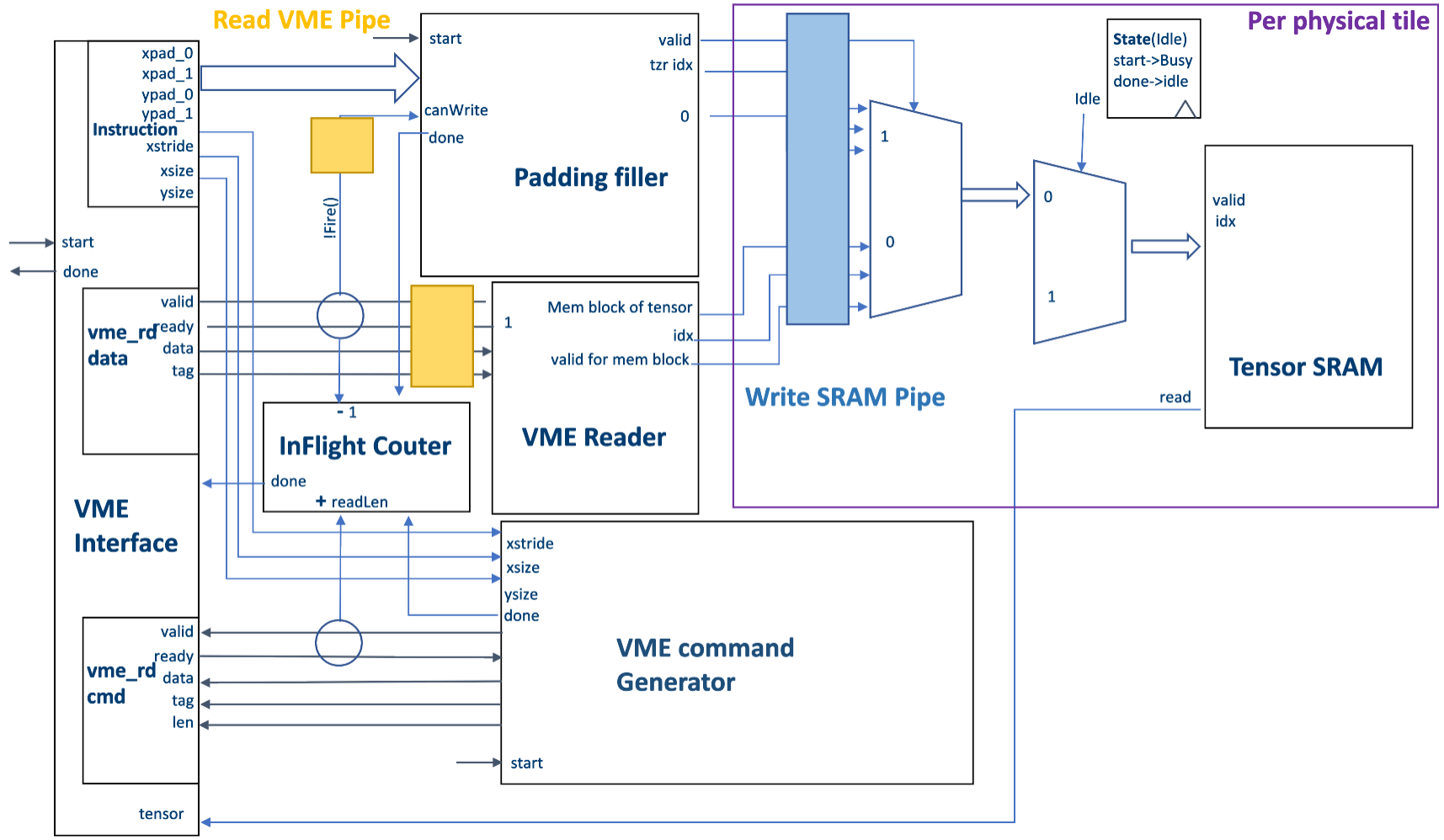}
  \caption{Separate Load CMD and VME Data transfer to process parallel load requests. Padding fill is executed when VME Reader is not busy.}
  \label{fig:memload}
\end{figure}

\subsubsection{Expanding the memory interface}

\begin{figure}
  \includegraphics[width=\linewidth]{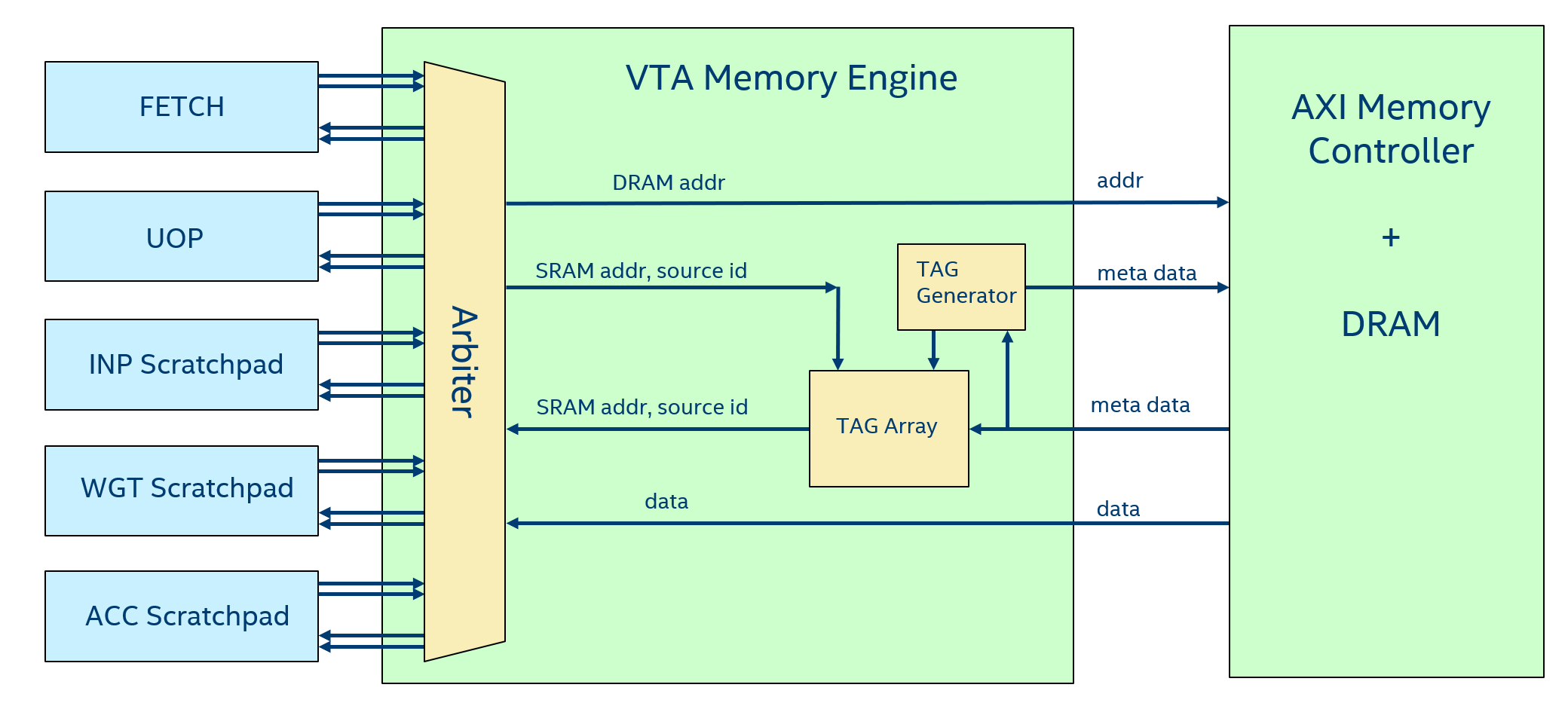}
  \caption{VTA Memory Engine enhanced with multiple outstanding requests and out-of-order completion. This micro-architecture allows multiple memory load requests to be inflight simultaneously. Complex metadata (scratchpad identifiers and addresses) are stored locally in the VME and only a tag is sent as metadata to the AXI memory controller. When the tag returns, it is used to recover the metadata and then send the load response data to the correct source scratchpad. The maximum number of inflight requests is limited by the size of this tag buffer and the capacity of the AXI memory controller.}
  \label{fig:vme}
\end{figure}

Existing VTA memory interface functionality is limited by 64-bit data interface. This limitation bounds computation by memory transfer. By expanding memory data interface to handle 64, 128, 256, 512-bit AXI data width, we achieved balanced compute-data transfer VTA performance for GEMM with up to 4K MAC operations.
TensorLoad functionality was generalized for all communications with VME. Two transfer modes were defined with respective implementations. Narrow data interface mode is used when AXI data width is less than destination data type width. Wide interface is for AXI data width wider than destination type width. For example, 128-bit AXI data loader into 32-bit UOP buffer uses wide loader implementation and can write 4 UOPs per cycle. 128-bit AXI data to 256 byte WGT scratchpad tensor loader uses a narrow type implementation with 16 AXI pulses per tensor load. The ratio of sizes between AXI and destination data should be power of 2. Instruction load is configured as 64/128/256/512 AXI data bits to 64-bit tensor loader with 128-bit tensor read interface. This allows to require 64-bit instruction address alignment instead of 128. Communication was separated into independent command generation and a data transfer interface communication  as shown in Figure \ref{fig:memload}. The number of inflight AXI data load requests is controlled by VME module.
VTA core TensorLoad module issues AXI data load requests which are queued by VME. Each AXI request is accompanied by a scratchpad destination address which is stored by VME together with request id. VME uses AXI id to identify request data destination and initiates data burst transfer, sends data and provides stored destination scratchpad address. A wide TensorLoad implementation also uses a destination address and a mask to mark valid bytes. The new functionality is summarized in Figure~\ref{fig:vme}.

\begin{figure}
  \includegraphics[width=\linewidth]{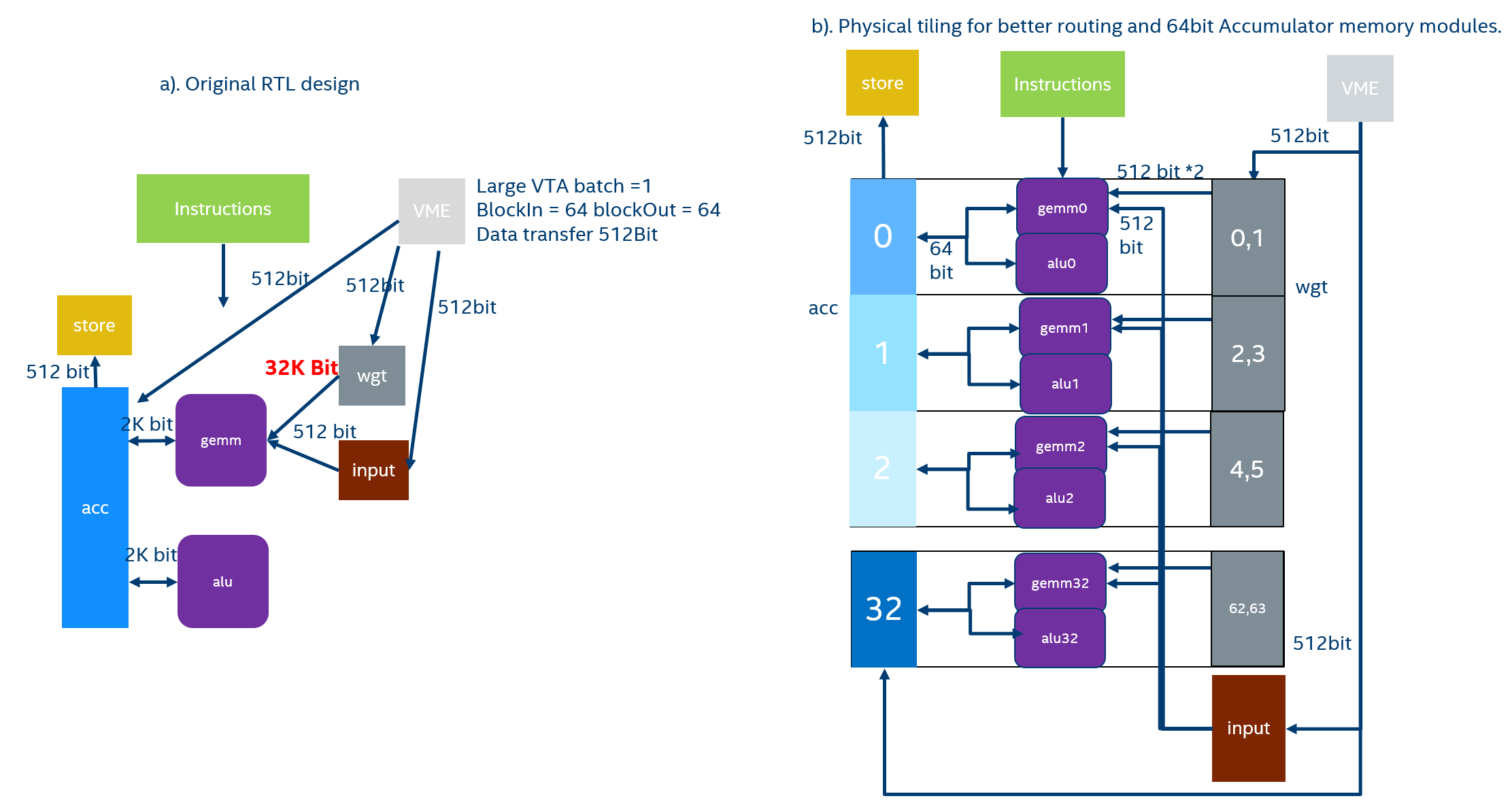}
  \caption{a). Original RTL design. All similar functionality is grouped leading to 32K bit data read interface. b). Tile is grouped around minimum GEMM functionality. TensorLoad is spread over whole die area.}
  \label{fig:physhier}
\end{figure}
\begin{figure}
  \includegraphics[width=\linewidth]{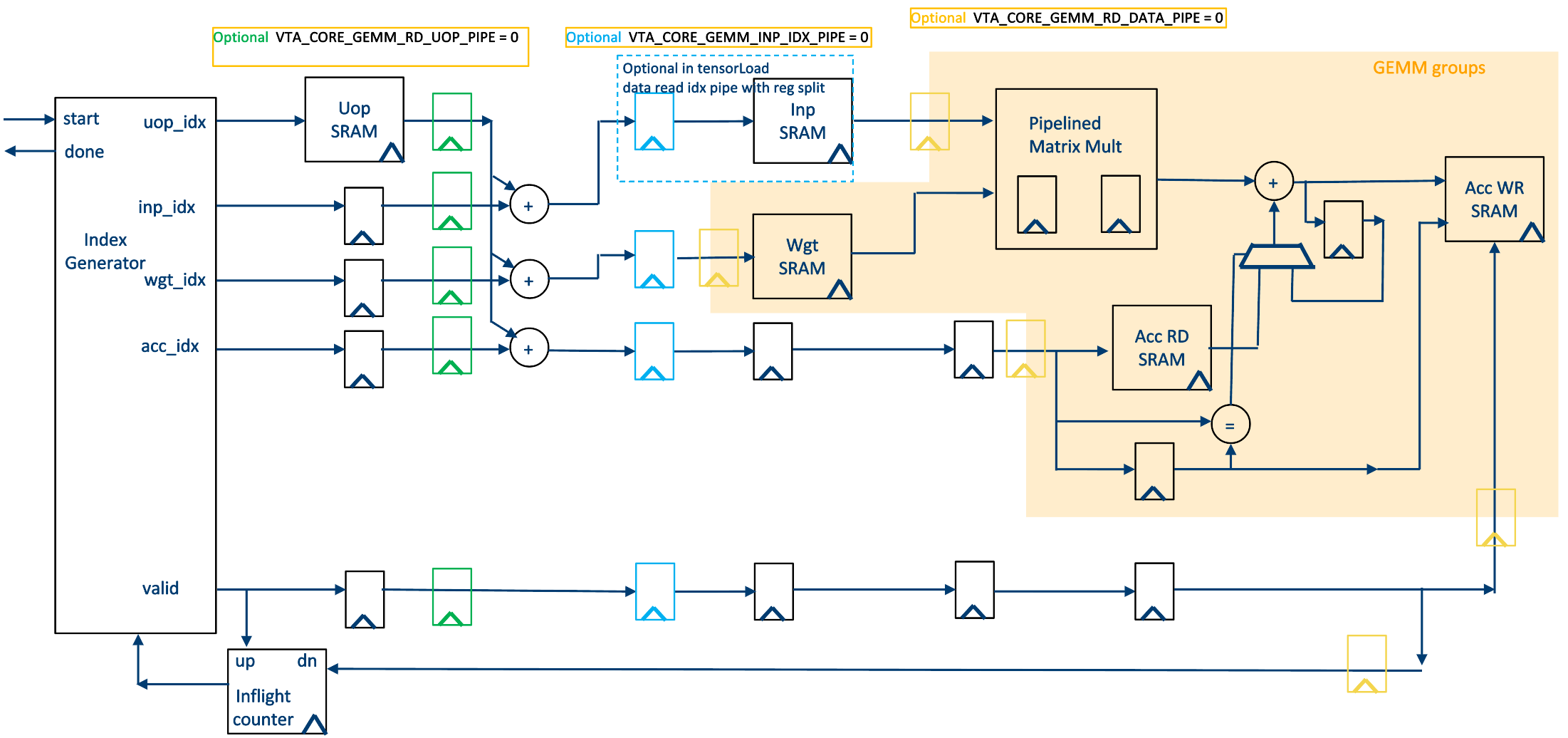}
  \caption{Distance from instruction decoder and to Input and Store scratchpads may require pipe stages for better timing.}
  \label{fig:pipewiregemm}
\end{figure}
\begin{figure}
  \includegraphics[width=\linewidth]{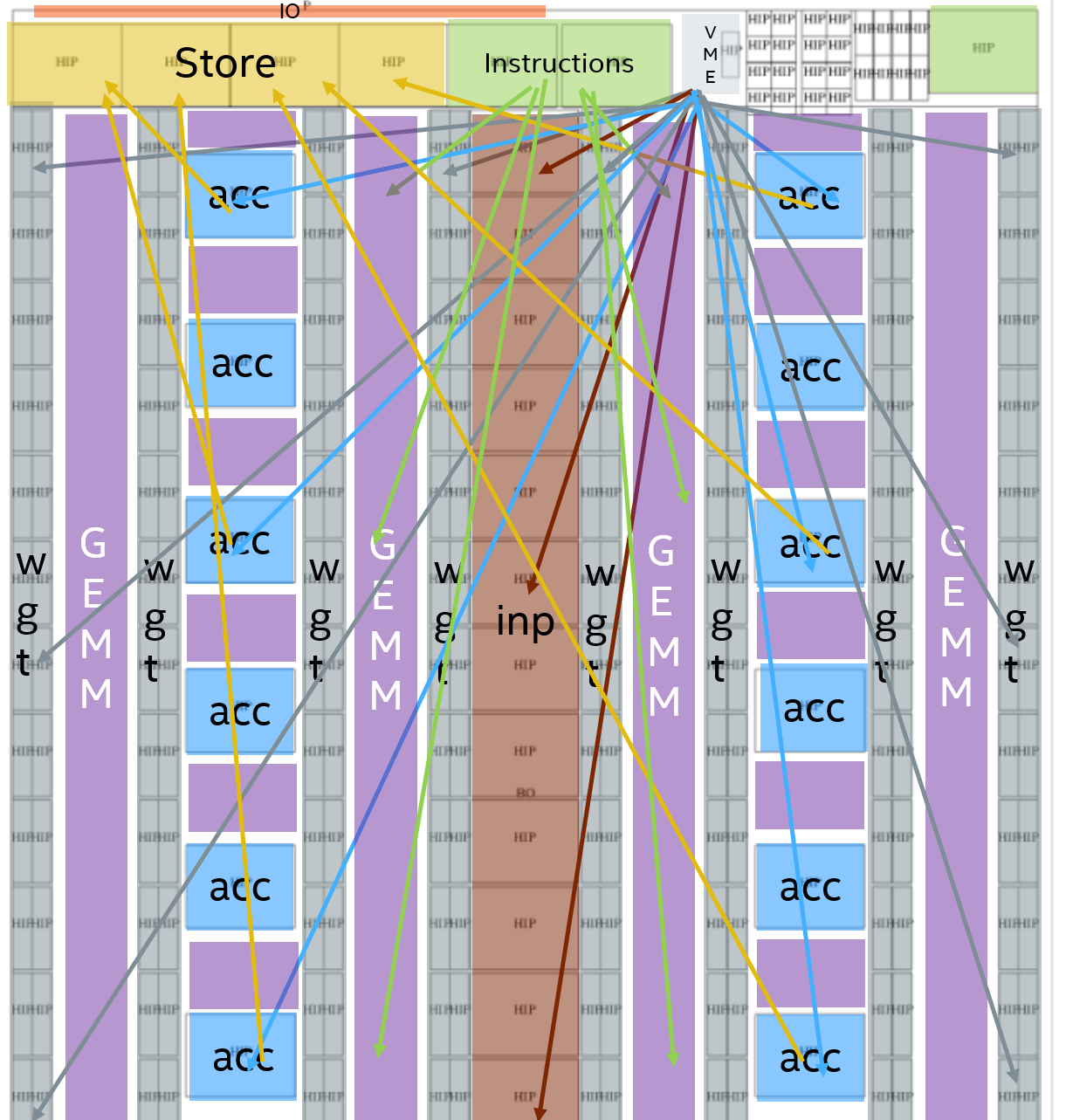}
  \caption{Weight, Input, Accumulator scratchpad memory modules are not grouped together. Large TensorLoad data transfer distance requires pipe stages.}
  \label{fig:pipewireload}
\end{figure}

\subsection{Flexible Floorplans}
A floorplanning Python library was created to support generation of multiple floorplans for different VTA configurations. Floorplan generator allows definition of layout objects with design sub-hierarchy name, width, height, and orientation. At bottom leaf level of floorplan hierarchy, we have macro modules which are usually instances of memory modules. They have width and height provided by tech description. Design sub-hierarchy can have placement bounds. Each floorplan has a definition in a form of a Python class hierarchy and implementation generated by running that program. Floorplanner provides capability to instantiate arrays of floorplan instances and flip individual objects if required. Result visualization and overlap/spacing, unique instance name checks are also provided.
The original VTA design hierarchy is built around GEMM/ALU/INP/WGT/ACC/OUT objects communicating with each other. Physical size of those object should be considered if VTA configuration is large enough for communication lines to require buffering. Another problem is a routing congestion. Wide WGT and ACC data buses produce too much congestion if floorplan placement bounds are based on original design hierarchy.
Physical design tools use design instances to define physical constraints. Multiple runs of physical flow are required to converge on timing and routing congestion metrics. Floorplanner allows to write a flexible program to generate multiple hierarchies of different size and orientation of objects with reuse of existing sub-hierarchies.
A better design hierarchy built around ACC scratchpad value was found as shown in Figure \ref{fig:physhier}. A center of this structure is an ACC memory module which is used to store GEMM/ALU result. GEMM/ALU read and write this module, so their logic is placed nearby, and it is timing critical and not being spread too far. Tested VTA configurations have BATCH of 1 or 2 which BLOCK$\_$IN and BLOCK$\_$OUT can be quite large up to 64. It means that WGT scratchpad memory bit width can be quite large and only a portion of it is used to get each ACC value. In a case of BATCH 1,  BLOCK$\_$IN 64 and BLOCK$\_$OUT of 64 it means 512 bit of WGT scratchpad are used to compute one and only one value of ACC. All 512 bits from INP scratchpad are used to get 64 different values of ACC. It makes sense to place a portion of WGT scratchpad close to respective ACC module. Not much could be done to distribute INP scratchpad. OUT scratchpad writing can be pipelined and that scratchpad can be placed anywhere depending on available space. The same UOP and instruction memory data is used in each GEMM/ALU and can be placed anywhere and buffered/pipleined. Major modules of design cannot be placed as separate instances as GEMM/ALU/ACC/WGT/INP content overlap each other all over die area.
\subsection{Wire Pipelining}
\subsubsection {GEMM pipelining}
GEMM and ALU have functionality which is common for every value to be calculated and a unique functionality which is specific for each result value as shown in Figure \ref{fig:pipewiregemm}. The result of common part calculation is delivered to each memory module and it is buffered and pipelined as read UOP and generate index pipes. WGT reading part of GEMM/ALU compute and ACC writing was grouped around ACC value and optimized for minimum delay and pipe stages. GEMM/ALU units were grouped into groups with as many units as needed to complete computation in one cycle while they all are fed by UOPs/Instruction/INP data from some distant source.
\subsubsection {Scratchpad load pipelining}
Scratchpad loading is a delivery of VME data or scratchpad tensor size number of bits to a respective memory module as shown in Figure \ref{fig:pipewireload}. In many VTA configurations WGT/ACC scratchpads are distributed all over die area and INP scratchpad is quite large. Data from VME cannot be delivered to them in one cycle and requires construction of a buffering tree. The original design hierarchy was changed and WGT/ACC/INP were split into groups. WGT/ACC groups are based on ACC value calculation code and how much can be computed at one logic stage.

\subsection{Compiler Modifications}

We modified the compiler for lowering all deep learning layers to different configurations of VTA with high performance. We introduce a heuristic-guided analytical scheme called \textit{Tiling Parameter Search (TPS)} for mapping deep learning operators such as standard convolution and depthwise convolution to different VTA configurations. We modify the TVM IR pass related to threading for improving double buffering of the GEMM and LOAD operations in VTA. In addition, we enable the execution of depthwise convolution and dense layers in VTA, thus providing full support for ResNet and MobileNet workloads in VTA.

\subsubsection{Tiling Parameter Search}
\label{subsec:tps}

The TVM compiler creates \textit{schedules} for representing tensor operators as a blueprint for lowering the computation to VTA through tensorize intrinsics and memory load/store instructions. The convolution schedule is expressed in a fixed scheduling template with a sequence of program transformation primitives such as loop reordering, tiling and threading. Different choices of the tiling parameters modify the data access patterns of the schedule and alters the ordering of memory accesses and compute unit execution, thereby directly affecting the performance. AutoTVM \cite{chen2019learning} and Ansor \cite{ansor2020} provide an automated search process to optimize program schedules. Both these schemes use a cost model that requires collecting performance metrics of candidate schedules from a hardware platform. During the VTA design enhancement, we required fast discovery of high-performance convolution implementations on different VTA configurations, for which we used TPS.

\begin{figure}
\centering
\includegraphics[width=0.9\linewidth]{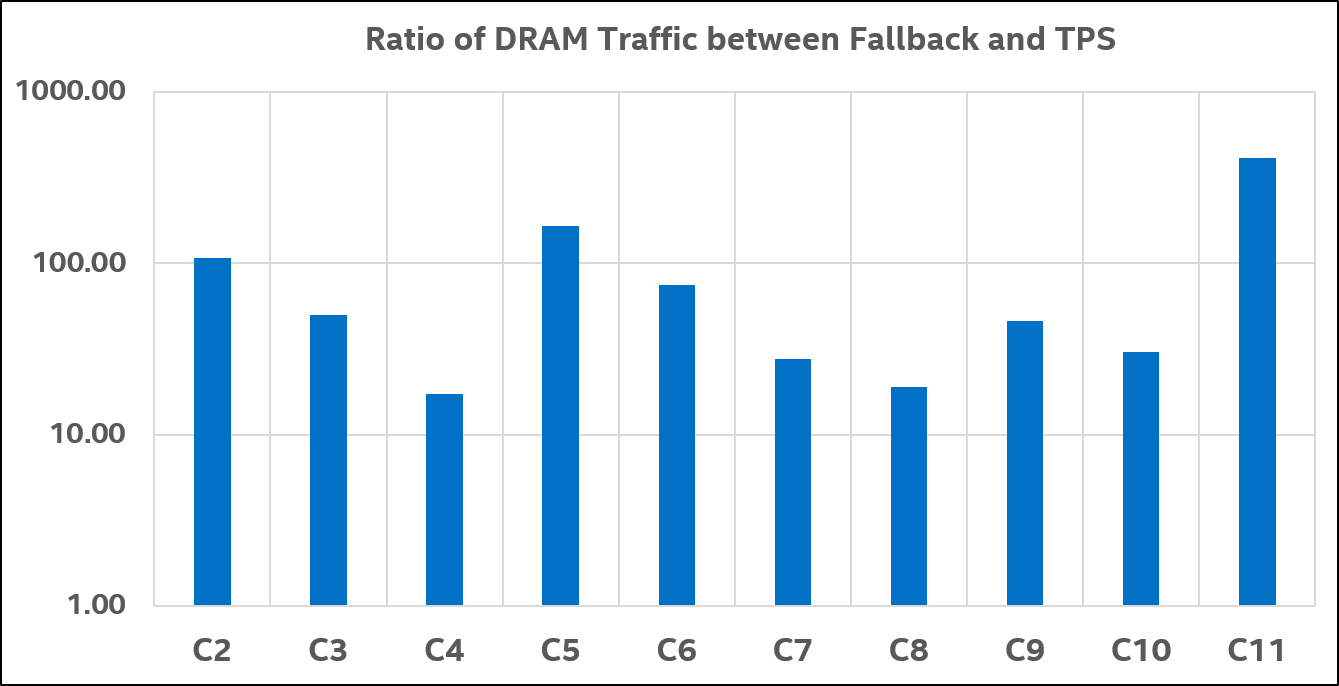}
\caption{With TPS, DRAM byte transfer is reduced by 20x-400x for different convolution layers on BLOCK=32 configuration}
\label{fig:tpsdram}
\end{figure}

For a given convolution schedule and a VTA configuration, we express the bytes transferred from DRAM to scratchpads as an analytical cost function of the tiling parameters, as expressed in Appendix \ref{app:tps}. We use this cost function and scratchpad size constraints to rapidly explore the tiling parameter space and optimize a convolution schedule for a given VTA configuration. Without the TPS algorithm, the default behavior of the TVM-VTA stack is to use a fallback schedule for convolution that guarantees compilability of a deep learning workload on any VTA configuration by ensuring minimal use of local scratchpad at the expense of high DRAM byte transfer. Using the TPS algorithm, we are able to reduce the DRAM byte transfer by 2-3 orders of magnitude as shown in Figure \ref{fig:tpsdram}. We plot the ratio of DRAM traffic (in bytes) created by the fallback schedule and the TPS-generated schedule for ResNet-18 convolution layers C2-C11 on a VTA configuration with BLOCK$\_$IN=BLOCK$\_$OUT=32.

\subsubsection{Double Buffering}

The double buffering implementation of VTA divides each of the input, weight and accumulator scratchpads into two different logical chunks for separate load and compute access. While one half of the scratchpads is loaded with data from DRAM, the compute units (GEMM and ALU) operate on data present in the other half. The existing implementation of the TVM compiler's virtual threading pass fetches redundant data from DRAM to the two halves of the scratchpad, thus failing to exploit data reuse. For example, with two halves of input and weight scratchpads represented as $I_1$, $I_2$ and $W_1$, $W_2$, two chunks of input and weight data represented as $d_{i1}$, $d_{i2}$ and $d_{w1}$, $d_{w2}$, the compiler loads $d_{i1}$ and $d_{w1}$ into $I_1$ and $W_1$ respectively. After the load completes, the GEMM module works on $I_1$ and $W_1$, whereas the load module loads $d_{i1}$ and $d_{w2}$ into $I_2$ and $W_2$. This causes redundant load of $d_{i1}$ twice. We modified the thread injection IR pass to automatically identify the redundant loads in alternative memory load threads and reuse the $d_{i1}$ for the second GEMM by modifying the uop access pattern sequence from $(I_1, W_1), (I_2, W_2), (I_1, W_1), (I_2, W_2)$ to $(I_1, W_1), (I_1, W_2), (I_2, W_1), (I_2, W_2)$. This modification reduces power consumption by suppressing redundant memory loads and for memory-bound workloads, this improves throughput as well.

\begin{figure}
\includegraphics[width=0.9\linewidth]{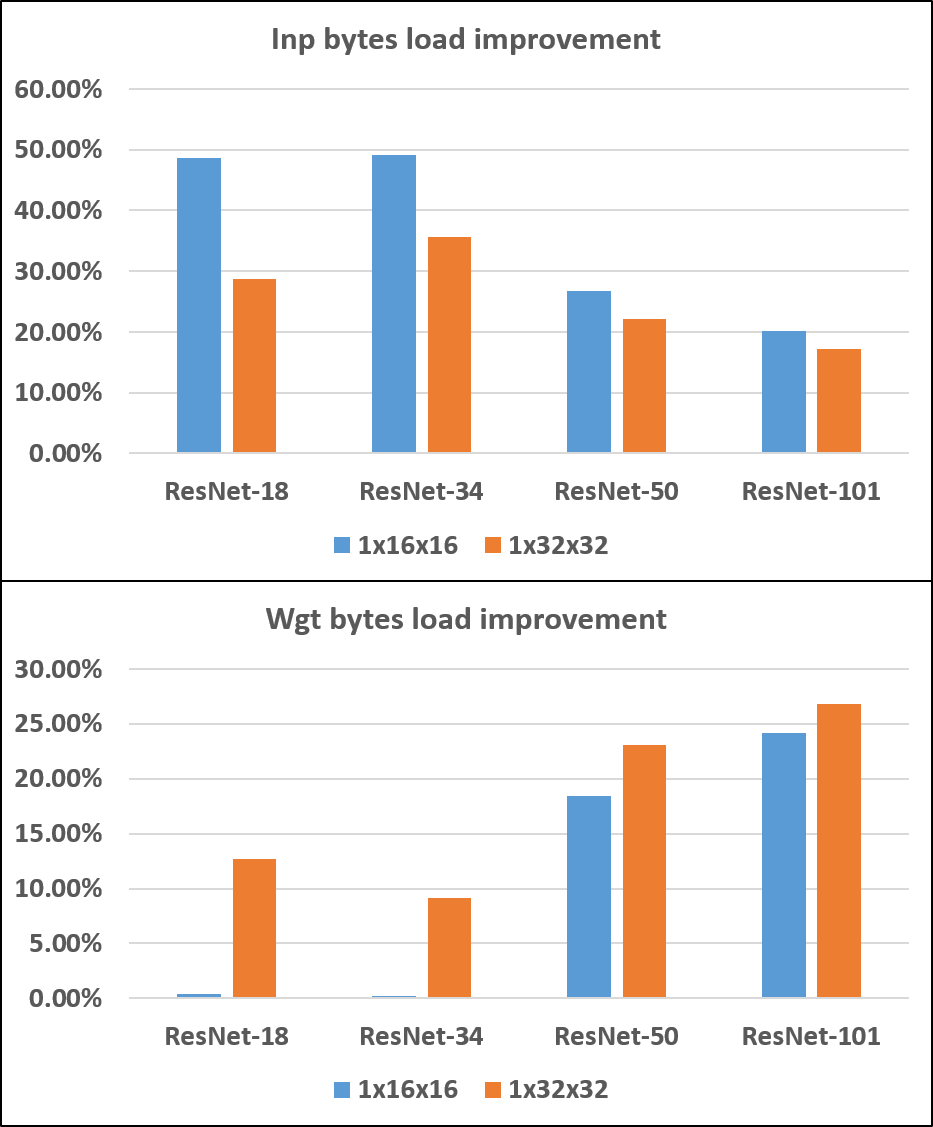}
\centering
\caption{Reduction in bytes loaded from DRAM to scratchpad (1x16x16 config implies VTA batch=1, block-in=16, block-out=16)}
\label{fig:doublebuffload}
\end{figure}

Figure \ref{fig:doublebuffload} shows the reduction in byte transfer from DRAM to input and weight scratchpads for 4 different ResNets on 2 VTA configurations. For a given workload on a particular VTA configuration, the total reduction of bytes loaded from DRAM is $\approx 50\%$, as the modification removes every other load of a data chunk. This, in turn, reduces the total power consumption by removing the redundant memory loads.

\begin{figure}
\includegraphics[width=0.9\linewidth]{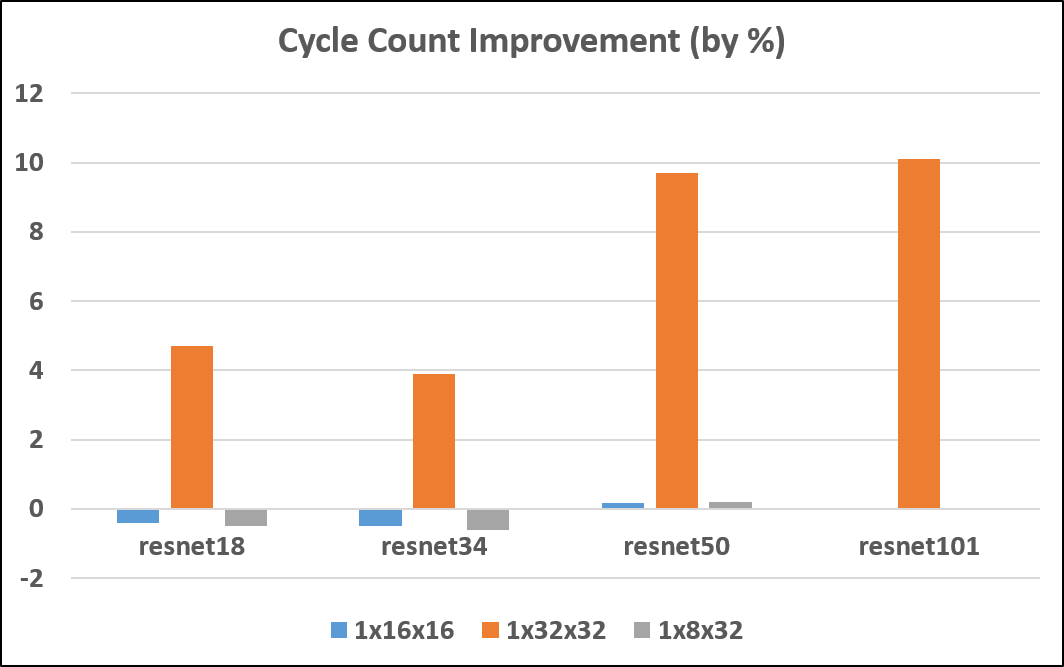}
\centering
\caption{Reduction in cycle count due to double buffering improvement}
\label{fig:doublebuffcycle}
\end{figure}

Figure \ref{fig:doublebuffcycle} shows the reduction in cycle count for 4 ResNets on 3 different VTA  configurations. For small networks (ResNet 18 and 34), the cycle count increases on small VTA configurations (256 MAC units) because of the higher uop memory loads. As these configurations are not memory-bound (and compute-bound instead), suppressing redundant DRAM loads doesn't improve throughput. For larger networks (ResNet 50 and 101) on compute-heavy configurations, reducing memory loads improves the performance by reducing the cycle count by $\approx 10\%$. 

\subsubsection{Depthwise Layers}
Depthwise separable convolution layers proposed in MobileNet\cite{mobilenet2017} architecture reduce the computational intensity of the convolution operation by factoring the convolution into depthwise and pointwise convolutions. The depthwise convolution applies a single filter to each individual channel, thus removing the channel-wide summation of standard convolution. This requires modifications in the TVM-VTA stack to support depthwise convolution layers. VTA's GEMM unit is a fused multiply-accumulate (MAC) unit that sums the partial product of input and weight over the input channels. As depthwise convolution doesn't sum the input-weight product over the input channels, we chose to utilize the VTA's ALU instead of the GEMM unit to execute depthwise convolution workloads. We created a new ALU opcode\cite{tvm_opensource, tvmvta_opensource} for element-wise multiplication and mapped a new schedule that divides up the depthwise computation into multiplications and additions. With this change, we are able to execute depthwise layers and in consequence, MobileNet network in VTA.

\subsection{End-to-end networks}
We created VTA schedules for average and max pooling layers by utilizing the ALU unit. With these modifications, we are able to execute the full ResNets from the 2nd convolution layer (1st convolution layer being channel-light at 3 channels is executed on the CPU by default \cite{moreau2018vta}) to the final fully-connected layer at the end. Similarly, we are able to execute the end-to-end MobileNet1.0 network with the added feature of executing depthwise layers on VTA. The capability of running end-to-end networks on the accelerator helps in targeted evaluation of accelerator design choices as we minimize the execution of DNN layers on CPU.

\subsection{Large design space}

One of the goals of our work was to expand the size of the design space. A
larger hardware design space obviously offers alternative points on the
performance-cost pareto curve. However, another benefit is that it may help
lower NRE costs if certain points have unanticipated synthesis or APR
challenges. Such benefits can only be realized if implemented throughout the
stack.

\begin{figure}
  \includegraphics[width=\linewidth]{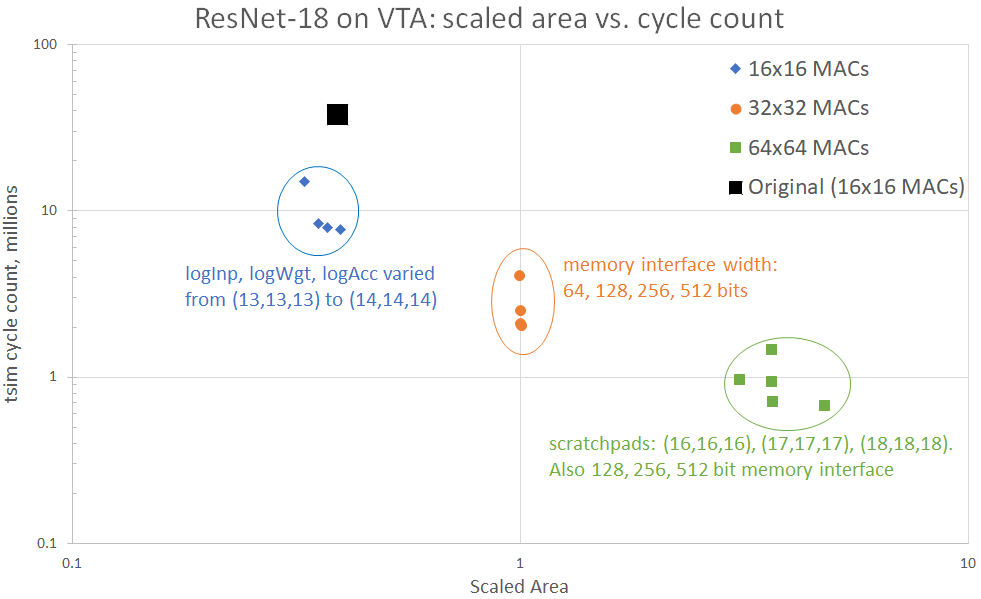}
  \caption{Cycle count vs. Scaled Area for a complete ResNet-18 Workload.
  The three ovals encapsulate points from 3 different MAC shapes: 4x4, 5x5,
  and 6x6. Within each oval, memory interface width and/or scratchpad sizes
  are varied.}
  \label{fig:designspace}
\end{figure}

The expanded design space we have enabled for TVM/VTA is summarized in
Figure \ref{fig:designspace}. The original TVM/VTA stack we started with
had a cycle count of 38 million cycles for this ResNet-18 workload, with
scaled area roughly equal to the other points with 4x4 MACs.

Overall, we can see that scaled area varies by an order of magnitude among
these points, while the cycle counts vary by a slightly greater range.
Scratchpad size is the main contributor to scaled area, while the cycle
count metric is affected by a variety of factors. Overall, a balancing of
the compute (captured by GEMM shape), memory bandwidth (captured by memory
interface width), and scratchpad size (captured by the addressable bits for
all scratchpads) is necessary to decrease the cycle count for this design.
Note that the relative impact of these factors also depends on the workload
being executed.

\subsection{Continuous Integration}

\begin{figure}
  \includegraphics[width=\linewidth]{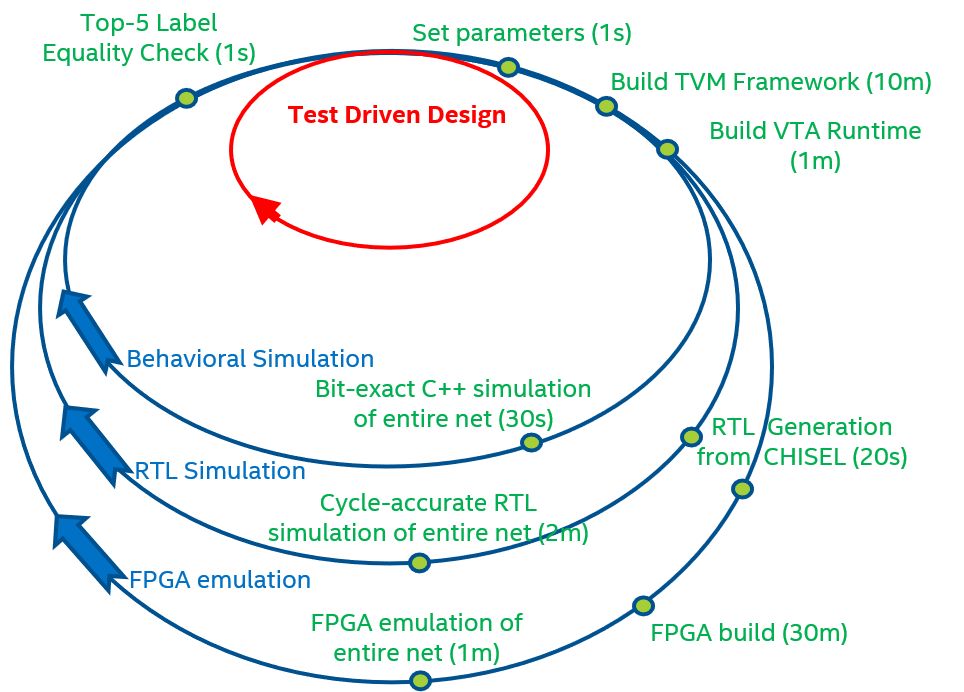}
  \caption{Continuous Integration quickly and automatically exercises cross-layer workloads.}
  \label{fig:ci}
\end{figure}

A key part of ensuring an always-alive model is to build a continuous integration (CI)
system which is capable of exercising cross-layer workloads on each commit. Figure \ref{fig:ci}
details the workloads, targets and approximate runtimes of the CI system we built for the
TVM/VTA stack. The 3 lower ovals show the 3 main targets which are exercised (\emph{fsim}, \emph{tsim},
and \emph{DE10 FPGA}) by cross-layer workloads. The small upper oval covers unit testing, which is
populated by CHISEL unit tests which are initially used for test-driven design.

The parameter setting, TVM framework build, runtime build, and equality checking stages are shared
amongst all targets. The common stage which consumes the largest runtime is building the TVM
framework, which takes around 10 minutes. However, TVM framework build is independent of VTA,
and we rarely made any modifications to it, meaning this stage could be skipped during most CI
runs.

We did not anticipate the high simulation speed we were able to reach with \emph{tsim}. Using the
verilator path, RTL simulation of VTA for several million cycles only took a few minutes.
End-to-end runs with accurate cycle counts were added to the CI pipeline for several
representative neural network workloads. This also allowed us to gauge the performance impact of
feature additions. Moreover, when backend runs were carried out later, we were able to
retroactively judge area vs. performance tradeoffs.

The role of the behavioral simulator \emph{fsim} changed as well. The main value of the
\emph{fsim} model was no longer its fast simulation speed, but instead its relative simplicity.
Functional discrepancies introduced by the CHISEL micro-architecture implementation in \emph{tsim}
were much easier to debug with the help of the \emph{fsim} behavioral reference model. Dynamic
trace-based validation capabilities were often used to find the earliest divergence between the
\emph{fsim} and \emph{tsim} models, lowering debug time.

The usage of the \emph{DE10 FPGA} target changed from our initial plan as well. Our early
investment into this platform was driven by the expectation that long traces were only
realistically possible on an FPGA platform, and that running such long traces would be a
frequent activity. However, the fast \emph{tsim} runtime confirmed the latter yet rejected
the former assumption. The relatively long FPGA build time was unable to be amortized among
the end-to-end runs since RTL simulation was fast enough for the relatively small number of
neural networks we targeted. In addition, multi-core RTL simulation was easily able to scale
unlike the device-limited FPGA scaling.

In our CI usage, the FPGA platform became more of an integration target:
were the assumptions regarding the environment made by the CHISEL model accurate? We were
able to find memory access timing bugs in the CHISEL implementation which were only exercised
in the context of a realistic FPGA platform and not in the context of the simpler environmental
model used by RTL simulation. In this debug, the dynamic trace instrumentation capabilities
were able to assist us to debug to a narrow window in time where the traces started to
diverge.

\subsection{Open source contributions}

The results described in this section have been open sourced in forked
repositories~\cite{tvm_opensource}~\cite{tvmvta_opensource}. Portions of this
work related to ALU and GEMM pipelining and memory bus width parameterization
have been merged into upstream tvm and tvm-vta repositories. These changes only
involve the CHISEL-generated-RTL (\textit{tsim}) target. Modifications to the
instruction set architecture to support larger field widths have not yet
been upstreamed since they affect additional targets. In addition, compiler
modifications needed to support additional layers and networks are not yet
upstreamed.

\section{Conclusions and Future Work}

Motivated by reducing the NRE cost for hardware accelerators, we decided to take a hardware/software co-design
approach to enhancing the TVM/VTA inference accelerator. Our methodology for enhancement was based on
incremental changes to features up and down the stack combined with always-alive modeling. We were able to
achieve significant increases in performance and vastly expand the size of the design space. Moreover, the
end-to-end workload evaluation time is low enough (a few minutes) that our enhanced stack can assist in
adjacent development activities, an example is exploration of the DNN architecture itself.

The design space exposed by our configurable inference accelerator can be jointly explored with the inference
workload to be run on the accelerator. Strategies to automatically discover better neural network
architectures~\cite{zoph2017neural} have emerged in the past few years. While initially focused on abstract
computation costs, optimization of inference latency for a given hardware target has
been explored in \cite{cai2018proxylessnas}. When such exploration is done at design time on a configurable
hardware target, a large set of additional options become available.

As we have seen, legal parameter values for our configurable inference accelerator have complex constraints.
Such constraints may be related to synthesis or place-and-route effects, available process-specific macros,
instruction set limitations, and compiler features. Resolving these issues to create a denser design space
may be possible with additional time, design effort, or increased hardware fabrication cost. The most expedient
design space, however, is likely sparse. This design space sparsity is ill-suited to current hardware-workload
co-optimization approaches. Based on this motivation, we have developed a pragmatic yet effective approach
based on realizable hardware in \cite{akhauri2021rhnas} to optimize both latency and energy-delay-product metrics.

\bibliographystyle{./IEEEtran}
\bibliography{tvm-vta-paper}

\appendix

\subsection{Tiling Parameter Search}
\label{app:tps}

A convolution layer is characterized by batch $b$, activation height $h$ and width $w$, kernel height $kh$ and $kw$, activation channels $fi$, output channels $fo$, padding height $ph$ and width $pw$, stride height $sh$ and width $sw$. The VTA GEMM unit is characterized by batch $b_{VTA}$, block-in factor $BI$ and block-out factor $BO$. The output height $oh$ and width $ow$ are calculated as
\begin{equation}
oh = \Bigl\lfloor\frac{h+2\times ph-kh}{sh}\Bigr\rfloor + 1, ow = \Bigl\lfloor\frac{w+2\times pw-kw}{sw}\Bigr\rfloor + 1
\end{equation}

For tensorizing the convolution to VTA GEMM intrinsics, the input and output channels are chunked up by block-in and block-out respectively, to create 2 additional dimensions in the tensor operations as $di=\frac{fi}{BI}$ and $do=\frac{fo}{BO}$. The scheduling template applies tilings to 5 different dimensions to create outer and inner tiling pairs. A dimension of size $n$ can be tiled as $(n_o, n_i)$, where $n_o\times n_i = n$. The 5 tiling dimensions are
\begin{equation*}
\frac{b}{b_{VTA}} \to (tb_o, tb_i),\; oh \to (th_o, th_i)
\end{equation*}
\begin{equation*}
ow \to (tw_o, tw_i),\; do \to (tco_o, tco_i)
\end{equation*}
\begin{equation*}
di \to (tci_o, tci_i)
\end{equation*}
In addition, two virtual thread dimensions are defined along output channel and input height dimension as $oc_n$ and $h_n$ that control the double buffering of weight or input scratchpad respectively. The virtual threading parameters can be either 1 or 2 to indicate whether double buffering is enabled or not. Both the values can't be simultaneously 2. The TPS algorithm is a constrained cost minimization problem, where the cost is the DRAM byte transfer and the constraints are the scratchpad usage. It can be expressed as 

\begin{equation}
\begin{aligned}
\min (l_{inp} + l_{wgt} + l_{acc})&\\
u_{inp}\ge 0 &\\
u_{wgt}\ge 0 &\\
u_{acc}\ge 0
\end{aligned}
\label{eq:min}
\end{equation}
where $l_{inp}$, $l_{wgt}$ and $l_{acc}$ are the bytes loaded into input, weight and accumulator scratchpads respectively. The under-utilization factors $u_{inp}$, $u_{wgt}$ and $u_{acc}$ ensure that a candidate tiling scheme is able to fit data fetches within the scratchpad sizes. Each of these factors are expressed through tiling parameters, workload parameters and VTA configuration parameters as,
\begin{equation*}
l_{inp} = tb_o\times \frac{th_o}{h_n}\times \frac{tco_o}{oc_n}\times tw_o\times tci_o\times s_{inp}
\end{equation*}
\begin{equation*}
l_{wgt} = tb_o\times \frac{th_o}{h_n}\times \frac{tco_o}{oc_n}\times tw_o\times tci_o\times s_{wgt}
\end{equation*}
\begin{equation*}
l_{acc} = tb_o\times th_o\times tw_o\times fo
\end{equation*}
where $s_{inp}$ and $s_{wgt}$ are the usages of the input and weight scratchpads and expressed as
\begin{equation}
\begin{aligned}
s_{inp} &= tb_i\times \frac{\frac{fi}{BI}}{tci_o}\times \left(\Bigl\lfloor\frac{\frac{h}{th_o}+2\times ph-kh}{sh}\Bigr\rfloor 
\times sh + kh\right)\\
& \times \left(\Bigl\lfloor\frac{\frac{w}{tw_o}+2\times pw-kw}{sw}\Bigr\rfloor\times sw + kw\right)\\
& \times b_{VTA}\times BI\times oc_n \times h_n
\end{aligned}
\end{equation}
\begin{equation}
\begin{aligned}
s_{wgt} &= \frac{\frac{fo}{BO}\times\frac{fi}{BI}\times kh\times kw\times BO\times BI}{tco_o\times tci_o}\times oc_n\times h_n
\end{aligned}
\end{equation}
The under-utilization factors of the scratchpads are calculated as the difference between the scratchpad size and the corresponding usage. With $c_{inp}$, $c_{wgt}$ and $c_{acc}$ as the capacity of the input, weight and accumulator scratchpads, the factors $u_{inp}$, $u_{wgt}$ and $u_{acc}$ are expressed as,
\begin{equation}
\begin{aligned}
u_{inp} &= c_{inp} - s_{inp}\\
u_{wgt} &= c_{wgt} - s_{wgt}\\
u_{acc} &= c_{acc} - s_{acc}
\end{aligned}
\end{equation}
with the accumulator usage $s_{acc}$ expressed as
\begin{equation}
\begin{aligned}
s_{acc} &= \left(\frac{\frac{b}{b_{VTA}}\times\frac{fo}{BO}\times oh\times ow\times b_{VTA}\times BO}{tb_o\times tco_o\times th_o\times tw_o}+\frac{fo\times b}{tb_o\times tco_o}\right)\\
& \times oc_n\times h_n
\end{aligned}
\end{equation}

The TPS algorithm exhaustively enumerates all the configurations in the tiling parameter space and finds the optimized schedule by evaluating the constrained cost minimization problem expressed in (\ref{eq:min}).

\end{document}